\documentclass[twocolumn]{aastex631}

\graphicspath{{./}{figures/}}
\usepackage{amsmath}
\usepackage{upgreek}
\usepackage[T1]{fontenc}
\usepackage{booktabs}

\usepackage{CJK}

\newcommand\sref[1]{\hyperref[#1]{\S~\ref*{#1}}}
\newcommand\fref[1]{\hyperref[#1]{Fig.~\ref*{#1}}}
\newcommand\Eqref[1]{Eq.~(\hyperref[#1]{\ref*{#1}})}
\newcommand\eeqref[1]{Eq.~\hyperref[#1]{\ref*{#1}}}
\newcommand\tref[1]{\hyperref[#1]{Table~\ref*{#1}}}
\newcommand\aref[1]{\hyperref[#1]{Appendix~\ref*{#1}}}

\newcommand{\Msun}{{\rm M_\odot}}

\shorttitle{Bridging scales: Modeling suppressed Bondi accretion on black holes and  its impact on galaxy growth}
\shortauthors{Porras-Valverde et al.}

\begin{document}
\begin{CJK}{UTF8}{mj}
%\UseRawInputEncoding
%\linenumbers
\title{Bridging scales: Modeling suppressed Bondi accretion on black holes and its impact on galaxy growth}

\author[0000-0002-1996-0445]{Antonio J. Porras-Valverde}
\affiliation{Department of Astronomy, Yale University, Kline Tower, 266 Whitney Avenue, New Haven, CT 06511, USA}
\email{antonio.porras@yale.edu}

\author[0000-0002-5554-8896]{Priyamvada Natarajan}
\affiliation{Department of Astronomy, Yale University, Kline Tower, 266 Whitney Avenue, New Haven, CT 06511, USA}
\affiliation{Department of Physics, Yale University, P.O. Box 208121, New Haven, CT 06520, USA}
\affiliation{Black Hole Initiative at Harvard University, 20 Garden Street, Cambridge, MA 02138, USA}

\email{priyamvada.natarajan@yale.edu}

\author[0000-0001-5287-0452]{Angelo Ricarte}
\affiliation{Black Hole Initiative at Harvard University, 20 Garden Street, Cambridge, MA 02138, USA}
\affiliation{Center for Astrophysics $\vert$ Harvard \& Smithsonian, 60 Garden Street, Cambridge, MA 02138, USA}
\email{angelo.ricarte@cfa.harvard.edu}

\author[0000-0003-1598-0083]{Kung-Yi Su}
\affiliation{Department of Physics \& Astronomy and Center for Interdisciplinary Exploration and Research in Astrophysics(CIERA), Northwestern University, 1800 Sherman Ave, Evanston, IL 60201, USA}
\affiliation{Black Hole Initiative at Harvard University, 20 Garden Street, Cambridge, MA 02138, USA}
\affiliation{Center for Astrophysics $\vert$ Harvard \& Smithsonian, 60 Garden Street, Cambridge, MA 02138, USA}
\email{kungyisu@gmail.com}

\author[0000-0002-2858-9481]{Hyerin Cho (조혜린)}
\affiliation{Center for Astrophysics $\vert$ Harvard \& Smithsonian, 60 Garden Street, Cambridge, MA 02138, USA}
\affiliation{Black Hole Initiative at Harvard University, 20 Garden Street, Cambridge, MA 02138, USA}

\author[0000-0002-1919-2730]{Ramesh Narayan}
\affiliation{Center for Astrophysics $\vert$ Harvard \& Smithsonian, 60 Garden Street, Cambridge, MA 02138, USA}
\affiliation{Black Hole Initiative at Harvard University, 20 Garden Street, Cambridge, MA 02138, USA}

\author[0000-0002-0393-7734]{Ben S. Prather} 
\affiliation{Black Hole Initiative at Harvard University, 20 Garden Street, Cambridge, MA 02138, USA}

\begin{abstract}\label{abstract}
The accretion and feedback processes governing supermassive black hole (SMBH) growth span an enormous range of spatial scales, from the Event Horizon to the circumgalactic medium. Recent general relativistic magnetohydrodynamic (GRMHD) simulations demonstrate that strong magnetic fields can substantially suppress Bondi accretion by creating magnetically arrested disk (MAD) states, reducing inflow rates by up to two orders of magnitude relative to classical Bondi predictions. We incorporate this magnetic suppression prescription from Cho et al. (2023, 2024) into the Dark Sage semi-analytic model (SAM), which tracks SMBH and galaxy co-evolution within hierarchical merger trees derived from the IllustrisTNG cosmological simulation. Implementing the suppression  across different Eddington-ratio regimes, we explore its impact on black hole mass functions (BHMFs), stellar mass functions (SMFs), and AGN luminosity functions. Restricting suppression to sub-Eddington accretors ($f_{\rm Edd} < 3 \times 10^{-3}$) and rescaling AGN feedback efficiencies gives simultaneous agreement with observed $z = 0$ SMFs and BHMFs, as illustrated by Case D in this work. At $z > 6$, super-Eddington growth episodes dominate in the SAM, reproducing JWST-inferred luminous AGN number densities. Our results highlight the critical sensitivity of galaxy assembly to the coupling between small-scale accretion physics and large-scale feedback regulation. Magnetic suppression of hot gas accretion can reconcile low-redshift constraints while preserving the rapid black hole growth required at early cosmic epochs, thereby providing a physically motivated bridge between horizon-scale GRMHD simulations and cosmological galaxy-formation models.
\end{abstract}

\section{Introduction}
\label{S:intro}
The growth of supermassive black holes (SMBHs) through gas accretion is governed by physical processes operating across extreme spatial scales, from horizon scales to galactic scales, where feedback processes regulate star formation. At the smallest scales, black hole's (BH) gravity dominates within a characteristic region known as the ``sphere of influence'', set by the Bondi radius $r_{\rm Bondi} = 2GM_{\rm BH}/c_s^2$, where the gravitational potential of the BH (determined by its mass $M_{\rm BH}$) balances the thermal pressure of the ambient gas (characterized by sound speed $c_s$), allowing accretion to proceed \citep{Bondi1952}. Although the Bondi accretion rate provides only an approximate description of gas capture within the ``sphere of influence,'' it offers a useful benchmark for comparing observations with theoretical models.

Recent observations of nearby low-luminosity AGN for instances when the Bondi radius is resolved find that the accretion rates are several orders of magnitude lower than conventional Bondi models for M87 \citep{Kuo2014, Russell2015, EventHorizonTelescopeCollaboration2021, Drew2025},  Sgr A* \citep{Quataert1999, Baganoff2003, EHTMW_5_2022}, and NGC 3115 \citep{Wong2011, Wong2014,Almeida2018}. The physical origin of this suppression has been uncertain until recently. Observations from GRAVITY and the Event Horizon Telescope reveal strong magnetic fields that thread the accretion flows in both Sgr A* and M87* \citep{GRAVITYColab_Abuter2018, GRAVITYColab_Jimenez-Rosales2020, EventHorizonTelescopeCollaboration2021, EHTMW8+2024}, suggesting that magnetic fields may regulate accretion. This hypothesis is now supported by recent simulations that bridge scales \citep{Ressler2018, XuWenrui2019, Ressler2020, Lalakos2022, Cho+2023, Cho+2024, Guo2024ApJ} across a wide dynamic range. In particular, the multizone method \citep{Cho+2023,Cho+2024,Cho2025} achieved a major breakthrough, for the first time allowing a single GRMHD simulation to capture both feeding and feedback, thereby achieving steady states over seven orders of magnitude in radius. These GRMHD simulations demonstrate that strong magnetic fields naturally produce magnetically arrested disk (MAD) states with accretion suppressed by two orders of magnitude compared to the conventional Bondi rate, and also transport black hole spin-dependent feedback power back to the galaxy \citep{Cho2025}.

While these simulations are idealistic in the sense that they track the growth history of a single SMBH, our goal in this work is to explore the impact of these findings for an ensemble of BHs in the Universe. In this work, we implement the Bondi accretion suppression prescription from \citet{Cho+2023, Cho+2024, Cho2025} for a population of SMBHs that accrete and evolve over cosmic time using the semi-analytic model (SAM) {\sc Dark Sage}. The key reason we choose {\sc Dark Sage} SAM to implement this prescription is its inclusion of a multiplicity of growth modes of BHs. In {\sc Dark Sage}, BHs grow via four distinct modes: (i) traditional hot gas accretion; (ii) cold gas merger-triggered accretion, which occurs sporadically; (iii) BH mergers, included in all SAMs and simulations with simplified assumptions regarding coalescence timescales; and (iv) cold gas disk instability-driven accretion, a novel process in {\sc Dark Sage}. The cold gas growth channels enable episodes of super-Eddington accretion. The availability of these distinct modes and associated feedback processes is a key feature of {\sc Dark Sage} that makes it uniquely suited to examine the mass assembly history of a population of SMBHs.

\section{Methodology}
\label{S:methods}

The semi-analytic model {\sc Dark Sage} tracks the angular momentum of gas and stars at different radii by evolving annuli with 30 equally-spaced logarithmic bins of fixed angular momentum \citep{Stevens2016}, a method inspired by \citet{Fu2010, Fu2013}. The basis for the model comes from {\sc SAGE} \citep{Croton2016}. To form a galaxy, we define a hot gas reservoir, which over a characteristic cooling time allows this gas to radiatively cool and condense \citep{White1978}. Once it cools down, the gas collapses to form a gravitationally-supported disk. {\sc Dark Sage} uses two annular disk structures, one for the stars and one for the gas, to calculate the rate of relevant physical processes such as star formation rates, disk instabilities, AGN and SNe energy dissipation, and metallicity gradients. In this work, we focus on improving the physicality of the SMBH prescriptions of {\sc Dark Sage} by adopting recently-developed formulae from GRMHD simulations \citep{Cho+2024,Cho2025}.  

We run {\sc Dark Sage} semianalytic model on dark matter only merger trees from the simulation {\sc IllustrisTNG100-1} \citep{2014Natur.509..177V, 2014MNRAS.444.1518V, 2014MNRAS.445..175G, 2015MNRAS.452..575S, Nelson2015}, which has a periodic box of length 110.7 Mpc with particle mass resolution of $8.9 \times 10^6\, \mathrm{M}_{\odot}$ and uses cosmological parameters derived from Planck \citep{2016A&A...594A..13P}, specifically $\Omega_M = 0.3089$, $\Omega_{\Lambda} = 0.6911$, $\Omega_b = 0.0486$, and $h = 0.6774$.

{\sc Dark Sage} seeds zero mass BHs for every halo in the underlying N-body simulation. At each time step, multiple channels simultaneously contribute to a SMBH's overall accretion rate:

\begin{itemize}
    \item Hot gas -- BH accretes hot gas coming from the hot gas reservoir (CGM) of its host galaxy.
    \item Cold Gas: Merger-triggered -- BH accretes a fixed fraction of cold gas from the ISM of the secondary galaxy when a galaxy merger occurs. 
    \item Cold Gas: Secular\footnote{\citet{PorrasValverde2025MassAssembly} use alternative terminology: secular instabilities are classified as \textit{in} situ accretion (unstable gas originating within the galaxy), while merger-driven instabilities are \textit{ex} situ accretion (gas acquired during mergers that becomes unstable).} -- BH accretes cold gas (from the ISM) that has preferentially low angular momentum. 
    \item BH-BH mergers: the model adds merging black holes together with no dynamical friction time delay. Mass loss during coalescence is not taken into account. 
\end{itemize}

\subsection{AGN accretion}
\subsubsection{Hot Gas Accretion}
\begin{center}
\noindent Parameters:  $\kappa_{\rm hot}$
\end{center}
We modeled hot gas accretion onto the central black hole at a rate approximated using the Bondi-Hoyle formula \citep{Bondi1952}:

\begin{equation}
\dot{M}_{\rm Bondi} = \pi G^2\, \frac{M_{\rm BH}^2\, \rho_0}{c_{\rm s}^3} \left[\frac{2}{5-3\gamma}\right]^{(5-3\gamma)/2(\gamma-1)},
\label{physicalBH}
\end{equation}
where the adiabatic index $\gamma=1.4$, the sound speed $c_{\rm s}$ is the thermal velocity scale of the ambient gas, estimated from the virial velocity of the host halo, $V_{\rm vir}$, and $\rho_0$ represents the density of the hot gas surrounding the black hole. \citet{Croton2006} estimates $\rho_0$ by relating the sound travel time across a shell with diameter twice the Bondi radius to the local cooling time, following the so-called ``maximal cooling flow'' model from \citet{Nulsen2000}. The cooling time for gas with local density $\rho_g(r)$ and temperature $T_{\rm vir}$ can be estimated by dividing its specific thermal energy by the cooling rate per unit volume:

\begin{equation}
t_{\rm cool} = \frac{3}{2} \frac{\bar{\mu}\, m_p\, k\,T_{\rm vir} }{ \rho_g(r)\, \Lambda(T_{\rm vir},Z)},
\label{tcool}
\end{equation}

\noindent
where $\bar{\mu}\, m_p\,$ is the mean particle mass, $k$ is the Boltzmann constant and $\Lambda (T,Z)$ is the cooling function, which depends on the gas temperature and its metallicity $Z$. Since $r_{\rm Bondi} \equiv 2G M_{\rm BH}/c_{\rm s}^2$, then:

\begin{equation}
t_{\rm cool} = \frac{4\, G\, M_{\rm BH}}{c_{\rm s}^3}~.
\label{tcool_bondi}
\end{equation}

Therefore, 

\begin{equation}
\rho_0 = \frac{3\, \bar{\mu}\, m_p\, k\,T_{\rm vir}\, c_{\rm s}^3 }{8\, G\, M_{\rm BH}\, \Lambda(T_{\rm vir},Z)}~.
\label{rho0}
\end{equation}
By substituting this density into Equation~\ref{physicalBH}, the accretion rate can be written as a function of the local temperature and black hole mass:

\begin{equation}
\dot{M}_{\rm BH, hot} = \kappa_{\rm hot} \ \frac{15}{16} \uppi\, G\, \bar{\mu}\, m_p\, \frac{k\,T_{\rm vir}\,}{\Lambda(T_{\rm vir},Z)\,} m_{\mathrm{BH}}~.
\label{eq:BondiAccretion}
\end{equation}
We denote $\dot{M}_{\rm BH, hot}$ as the accretion rate that is directly tied to the \textit{radio-mode} regime (see section \ref{radiomodeAGN}). The term $\kappa_{\rm hot}$ is the hot gas efficiency parameter referred to as $\kappa_{\rm R}$ in \citet{Croton2016} that suppresses the accretion rate relative to a maximal cooling flow model. As a result, this is the factor by which the \textit{radio-mode} accretion is suppressed.

\subsubsection{Cold Gas Accretion: Mergers}

\begin{center}
\noindent Parameters:  $\kappa_{\rm cold, merger}$, $f_\mathrm{BH}$
\end{center}

When two dark matter halos merge, the galaxies merge and in {\sc Dark Sage} at this juncture the amount of cold gas funneled into the black hole as a result of a galaxy merger is given by  \citet{Kauffmann2000}:

\begin{multline}
M_{\rm BH, cold, merger} = \kappa_{\rm cold, merger} * f_{\rm BH} \left[1 + \left( \frac{280\,\mathrm{km\,s}^{-1}}{V_{\rm vir}}\right)^2 \right]^{-1} \\ \sum_{i=1}^{30} (m_{i, \rm cen} + m_{i, \rm sat})\, \mathrm{min} \left( \frac{m_{i, \rm sat}}{m_{i, \rm cen}}, \frac{m_{i, \rm cen}}{m_{i, \rm sat}} \right)~,
\label{eq:coldmodeaccretion}
\end{multline}

\noindent
where $\kappa_{\rm cold, merger}$ accounts for any suppression associated with cold gas merger-triggered accretion\footnote{Note that $\kappa_{\rm cold, merger}$ is defined differently in this work compared to \citet{Croton2016} and \citet{Stevens2016}. We introduce it here in Equation~\ref{eq:coldmodeaccretion} as a suppression factor of the accretion for consistency with our terminology. In contrast, \citet{Croton2016} and \citet{Stevens2016} use $\kappa_{\rm Q}$ to denote the \textit{quasar-mode} AGN feedback efficiency, which corresponds to $\epsilon_{\rm cold, merger}$ in our notation.}, $f_{\mathrm{BH}}$ represents the fraction of centrally funneled gas that goes to the BH instead of forming stars. The $m_i$ terms correspond to the amounts of gas contributed to the $i$-th annulus by the central and satellite galaxies. Since the angular momentum ($j$) distribution in galactic disks is skewed toward lower values, particularly for gas involved in mergers, the black hole tends to accrete preferentially low angular momentum gas. In the model, $f_{\mathrm{BH}}$ is a free parameter that is set to 0.3, and it represents the fixed fraction of gas that is accreted into the black hole. The remaining fraction, $1-f_{\rm BH}$, is deposited in the cold gas disk of the primary galaxy. We convert Equation~\ref{eq:coldmodeaccretion} to an accretion rate taking the difference in accreted mass between timesteps and dividing by the time interval. 

\begin{equation}
\dot{M}_{\rm BH, cold, merger} = \frac{M_{\rm BH, cold, merger} - M_{\rm BH, cold, merger, previous}}{{\delta t}}~.
\label{eq:Quasarmdot}
\end{equation}

\subsubsection{Cold Gas Accretion: Secular}

\begin{center}
\noindent Parameters: $f_{\rm move}$ 
\end{center}

During every timestep of galaxy evolution, {\sc Dark Sage} checks for Toomre instabilities \citep{Toomre1964}. If unstable gas is present, the corresponding annulus in the galactic disk triggers a starbust phase. Any leftover gas that is not resolved is transferred inward to its neighboring annulus. We denote $f_{\rm move}$ as the fraction of unstable gas that moves to an adjacent annulus. If the unresolved unstable gas reaches the innermost annulus, the gas is added to the central black hole. Analogously to Equation~\ref{eq:Quasarmdot}, we compute $\dot{M}_{\rm BH, cold, secular}$ from the change in the accreted mass between the time steps divided by the time interval. As instabilities can occur secularly or due to galaxy mergers, we separate these two contributions in our book-keeping. 

\subsubsection{Total BH accretion}

We define the total BH accretion rate as $\dot{M}_{\rm BH, in}=\dot{M}_{\rm BH, hot} + \dot{M}_{\rm BH, cold}$, where $\dot{M}_{\rm BH, cold} = \dot{M}_{\rm BH, cold, merger}+\dot{M}_{\rm BH, cold, secular}$. We refer to $\kappa_{\rm hot}$, $\kappa_{\rm cold, merger}$, and $\kappa_{\rm cold, secular}$ as the accretion efficiencies for hot gas, cold gas: merger-triggered, and cold gas: secular, respectively. Hereafter, $\kappa_{\rm cold}=\kappa_{\rm cold, merger} =\kappa_{\rm cold, secular}$. We then estimate the Eddington ratio:

\begin{align}
f_{\rm edd} &= \frac{\dot{M}_{\rm BH, in}}{\dot{M}_{\rm Edd}}, \label{eq:fedd}\\
\dot{M}_{\rm Edd} &= \frac{4\pi GM_{\rm BH}m_p}{\eta \sigma_T c}~,
\label{eq:mdot_Edd}
\end{align}
where $\sigma_T$ is the Thomson scattering cross-section and $\eta=0.1$ is the radiative efficiency. In {\sc Dark Sage}, only $\dot{M}_{\rm BH, hot}$ is capped at the Eddington limit, while $\dot{M}_{\rm BH, cold}$ can proceed at super-Eddington rates.

\subsection{AGN Feedback}

\begin{center}
\noindent Parameters:  $\eta$
\end{center}

{\sc Dark Sage}, like most SAMs, includes two distinct modes of AGN feedback: \textit{the radio mode} and \textit{the quasar mode}. Note that, as with mass accretion, both modes operate simultaneously.  \textit{The radio mode} is coupled to hot gas accretion. In this mode, the AGN injects energy into the CGM, heating the gas and potentially suppressing further star formation. \textit{The quasar mode}, meanwhile, is coupled to the cold gas that gets accreted when triggered by a merger. In this mode, the AGN heats up the ISM and can drive powerful outflows that can impact the entire host galaxy. While the \textit{quasar mode} is triggered by galaxy mergers, \textit{the radio mode} feedback continuously operates at every stage of the galaxy's mass assembly. Both feedback channels operate with efficiency $\eta=\dot{E}/\dot{M}_{\rm BH, in} c^2=0.1$.

\subsubsection{\textit{Radio Mode}}\label{radiomodeAGN}

\begin{center}
\noindent Parameters:  $\epsilon_{\rm R}$
\end{center}

\textit{Radio mode} feedback is proportional to the accretion rate of the hot gas.

\begin{equation}
\dot{E}_{\rm BH, R} = \epsilon_{\rm R}\ \eta \ \dot{M}_{\rm BH, hot} \ c^2 ~,
\label{eq:radio_mode_mheat}
\end{equation}

\noindent
where $\epsilon_{\rm R}$ is the coupling efficiency of the \textit{injection of radio-mode} energy into the CGM. As in \citet{Croton2016}, we assume that the cold gas is heated by the \textit{radio mode} feedback out to some radius ($r_{\mathrm{heat}}$) within which the gas can no longer cool and form stars \citep[see Equation 18 of][]{Croton2016}. The combined factor $\epsilon_{\rm R}\eta$ represents the \textit{net} radio-mode efficiency, set to the GRMHD informed value.

\subsubsection{Quasar Mode}
\begin{center}
\noindent Parameters: $\epsilon_Q$
\end{center}
{\sc Dark Sage} calculates the total energy produced by \textit{quasar mode} feedback using $M_{\rm BH, cold, merger}$:
\begin{equation}
\dot{E}_{\rm BH,Q} \times {\delta t} = \epsilon_{\rm Q}\, \frac{1}{2} \ \eta\, M_{\rm BH, cold, merger}\, c^2\, ~,
\label{eq:quasarmodefeedback}
\end{equation}

\noindent
where ${\delta t}$ is the timestep and $\epsilon_{\rm Q}$ is the efficiency with which the quasar wind causes the gas to either be transferred to the gas reservoir or escape the halo. To determine which path the gas would take, {\sc Dark Sage} calculates the total thermal energy of the cold and hot gas components:

\begin{equation}
\dot{E}_{\rm cold} \times \delta t = \frac{1}{2}\, M_{\rm cold}\, V^2_{\rm vir}\, \ {\rm and}\, \ \dot{E}_{\rm hot} \times \delta t = \frac{1}{2}\, M_{\rm hot}\, V^2_{\rm vir}\, ~.
\label{eq:quasar_mode_hotcoldenergy}
\end{equation}

\noindent
Thus, if the total energy, $E_{\rm BH,Q}$, is greater than $E_{\mathrm{cold}}$, the cold gas is added to the ejected gas reservoir within the halo. If the total energy is greater than $E_{\mathrm{cold}} + E_{\mathrm{hot}}$, then both cold and hot gas are ejected outside of the halo. {\sc Dark Sage} applies feedback energy to heat gas annuli sequentially from the center outward until depleted. Any excess energy subsequently ejects hot gas from the system.

\subsection{Growth Suppression from Magnetic Fields}\label{subsec: BHgrowth}

X-ray and radio observations of the M87* SMBH reveal accretion rates orders of magnitude below those expected from standard Bondi accretion\footnote{We note an important distinction: our Bondi accretion rate calculation includes only hot gas, following the traditional semi-analytic model approach, whereas cosmological simulations generally compute Bondi rates from both hot and cold gas.} prescriptions \citep{Wang2013, Russell2015, EventHorizonTelescopeCollaboration2021, Drew2025}. GRMHD simulations find that in the hot accretion regime, much of the inflowing gas is prevented from reaching the black hole by the braking effect of magnetic fields \citep{,Ressler2018, Ressler2020, Lalakos2022, Cho+2023,Cho+2024, Guo2024ApJ, Cho2025, Guo2025ApJ}.\citet{Cho+2024} provides an empirical formula, all based on GRMHD simulations, for the accretion suppression factor, $\kappa_{\rm cho}$, by which the accretion rate on the BH is suppressed relative to the classic hydrodynamic Bondi accretion rate:
\begin{align}
    \kappa_{\rm cho}(T_{\rm vir}) &= \left(\frac{r_{\rm Bondi}(T_{\rm vir})}{12\,r_g}\right)^{-0.5}\label{eq:kappa_cho_1} \\
    &= \sqrt{\frac{6\gamma k T_{\rm vir}}{\mu m_p c^2}},
\end{align}
where $r_g \equiv G M_{\rm BH}/c^2$ is the gravitational radius.\footnote{Note that the definition of Bondi radius $r_{\rm Bondi}$ is a factor of 2 larger here than in \citet{Cho+2024}, leading to a different numerical factor (12 instead of 6) in \autoref{eq:kappa_cho_1}. A similar formula was derived in MHD simulations by \citet{Guo2024ApJ}}
We explore the effect of implementing \citet{Cho+2024}'s BH suppression factor $\kappa_{\rm cho}$ in the semi-analytic model {\sc Dark Sage}. The power of {\sc Dark Sage} rests in the ability to reproduce well the empirically determined local stellar mass function (SMF) and the black hole mass function (BHMF). To this end, we explore and compare the following four scenarios: 

\begin{itemize}
    \item {\bf Case A:} the baseline fiducial model {\sc Dark Sage} that best matches the current observations of the SMF and BHMF at $z=0$.
    \item {\bf Case B:} implementation of $\kappa_{\rm cho}$ for accretion across all Eddington ratios and therefore suppression of all growth channels available for BH growth in {\sc Dark Sage}. ($\kappa_{\rm hot} =\kappa_{\rm cold}=\kappa_{\rm cho} $)
    \item {\bf Case C:} implementation of $\kappa_{\rm cho}$ only for BHs accreting at extremely low rates, with Eddington ratios
    $f_{\rm Edd}<{3 \times 10^{-3}}$ - this is the scenario corresponding to the hot accretion flows reported in \cite{Cho+2023,Cho+2024}. For  $f_{\rm Edd}>{3 \times 10^{-3}}$, we set $\kappa_{\rm hot}=\kappa_{\rm cold}=1$. 
    \item {\bf Case D:} same as {\bf Case C} but changing the efficiency of the AGN feedback coupling ($\epsilon_R$ and $\epsilon_Q$), recalibrating it so that the total mass of gas heated by \textit{radio-mode} and \textit{quasar-mode} feedback is more similar to {\bf Case A}.
\end{itemize}

In {\bf Case B}, we explore the impact of implementing the suppression factor, $\kappa_{\rm cho}$, to all central and satellite galaxies that host central BHs across all growth channels available in {\sc Dark Sage}. In {\bf Case C}, we adopt a more physical approach explored in GRMHD simulations. We apply the suppression factor only when the Eddington ratio (calculated before suppression) falls below $f_{\rm Edd}<{3 \times 10^{-3}}$, corresponding to the transition from inefficient hot flows to more efficient thin-disk accretion at higher Eddington ratios. {\bf Case D} is similar to {\bf Case C} however, here we assume that not all of the energy liberated by accretion couples to the galaxy. As energy is radiated outward, only a fraction of this energy efficiently couples with and heats the surrounding gas to produce effective feedback. Therefore, we rescale the \textit{radio-mode} and \textit{quasar-mode} coupling efficiencies via:

\begin{equation}
  \epsilon_{\rm rescale} = \mathrm{min}\left( \ \frac{3\times10^{-3}}{\kappa_{\rm hot}}, 1 \ \right) ~.
\label{eq:quasar_mode_accre_eff}
\end{equation}

\noindent For $\kappa_{\rm hot} > 3\times 10^{-3}$, the coupling efficiencies, $\epsilon_{\rm R}$ and $\epsilon_{\rm Q}$, are increased to produce an increased amount of feedback required to quench massive galaxies. For {\bf Case D}, the rescaling factor $\epsilon_{\rm rescale}$ depends on the Eddington ratio regime. At low Eddington ratios ($f_{\rm Edd}<{3 \times 10^{-3}}$), BH accretion is suppressed with $\kappa_{\rm hot} =\kappa_{\rm cold}= \kappa_{\rm cho}$, yielding $\epsilon_{\rm rescale}\sim 0.9-1$ depending on the specific value of $\kappa_{\rm cho}$ (see Figure~\ref{fig:kappachohist_BHbins_z0_z6}). At high Eddington ratios ($f_{\rm Edd}>{3 \times 10^{-3}}$), no accretion suppression is applied ($\kappa_{\rm hot} =\kappa_{\rm cold}= 1$), so $\epsilon_{\rm rescale}\sim 3 \times 10^{-3}$. 

In general, the stellar mass and black hole mass functions are more sensitive to rescaling changes in $\epsilon_{\rm R}$ than they are to those in $\epsilon_{\rm Q}$. This is because \textit{radio-mode} feedback operates continuously throughout the evolutionary history of galaxies and the BHs they host, whereas the \textit{quasar-mode} feedback is triggered only episodically during galaxy mergers.

\begin{table*}
\centering
%\begin{tabular*}{\textwidth}{@{\extracolsep{\stretch{1}}}l l l p{5cm}@{}}
%\begin{tabular}{lllp{5cm}}
\begin{tabular}{cccc}
\toprule
Model name & BH suppression details & BH accretion & AGN feedback \\
\midrule
{\bf Case A} & - & $\kappa_{\rm hot}=3\times10^{-3}$ & $\eta = 0.1, \ \epsilon_{\rm R}=1$  \\
Fiducial & & $\kappa_{\rm cold}=1$ & $\epsilon_{\rm Q}=7\times10^{-4}$ \\
\midrule
{\bf Case B} & Suppressing all BH & $\kappa_{\rm hot}=\kappa_{\rm cho}$ & $\eta = 0.1, \ \epsilon_{\rm R}=1$ \\
& growth channels & $\kappa_{\rm cold}=\kappa_{\rm cho}$ & $\epsilon_{\rm Q}=1$ \\
\midrule
{\bf Case C} & Suppressing BH growth & For $f_{\rm Edd}<3\times 10^{-3}$:  & $\eta = 0.1$ \\
& for BHs accreting at & $\kappa_{\rm hot}=\kappa_{\rm cold}=\kappa_{\rm cho}$  & $\epsilon_{\rm R}=\epsilon_{\rm Q}=1$ \\
 & sub-Eddington & For $f_{\rm Edd}>3\times 10^{-3}$: & \\
 &  & $\kappa_{\rm hot}=\kappa_{\rm cold}=1$ & $\epsilon_{\rm R}=\epsilon_{\rm Q}=1$ \\
 
\midrule
{\bf Case D} & Suppressing BH growth for BHs& For $f_{\rm Edd}<3\times 10^{-3}$: & $\eta = 0.1$\\  & accreting at sub-Eddington & $\kappa_{\rm hot}=\kappa_{\rm cold}=\kappa_{\rm cho}$ & $\epsilon_{\rm R}=\epsilon_{\rm Q}=\epsilon_{\rm rescale}$ \\
& and scaling down feedback 
&For $f_{\rm Edd}>3\times 10^{-3}$:& \\
 & otherwise
 & $\kappa_{\rm hot}=\kappa_{\rm cold}=1$ & $\epsilon_{\rm R}=\epsilon_{\rm Q}=\epsilon_{\rm rescale}$ \\

\bottomrule
\end{tabular}
\caption{Detailed overview of the suppression schemes adopted in our Black Hole Growth Models. We denote accretion efficiencies as $\kappa_{\rm hot}$ (hot gas) and $\kappa_{\rm cold}$ (cold gas from merger and secular channels) in the BH accretion column and feedback coupling efficiencies as $\epsilon_{\rm R}$ and $\epsilon_{\rm Q}$ in the AGN feedback column. For clarity, we use $\eta$ to represent the overall feedback efficiency.}
\label{Table: sims_details}
\end{table*}

Table \ref{Table: sims_details} shows a summary of our different BH suppression models. In all of our simulations, we fix $f_{\rm BH}=0.3$\footnote{Given that $0\leq f_{\rm BH} \leq 1$, $\kappa_{\rm cho}$ suppression cannot be compensated by re-tuning $f_{\rm BH}$ by 1/$\kappa_{\rm cho}$.} and $\eta=0.1$, modeling feedback from BHs with intermediate spin $a_*\sim0.7$. \citep{Cho2025} find a slightly higher BH feedback ($\eta\sim0.3$) for the BH spin $a_*=0.9$ case.  Note that \textit{radio-mode} and \textit{quasar-mode} feedback are directly coupled to hot gas accretion and cold gas accretion respectively; therefore, any suppression applied on the accretion rate directly affects the corresponding feedback scale.

Our {\bf Case A} does not suppress cold gas accretion. It only suppresses the amount of \textit{quasar-mode} feedback using the coupling efficiency ($\epsilon_{\rm Q}$). Both $\kappa_{\rm hot}$ and $\epsilon_{\rm Q}$ are two out of seven free parameters used to calibrate {\bf Case A} to the scaling relations. {\bf Case B} implements $\kappa_{\rm cho}$ suppression for all accretion modes. {\bf Case C} applies $\kappa_{\rm cho}$ suppression to all accretion rates only when $f_{\rm Edd}<{3 \times 10^{-3}}$. For $f_{\rm Edd}>{3 \times 10^{-3}}$, BH accretion operates without suppression. Finally, {\bf Case D} employs the same accretion suppression criteria as {\bf Case C}, but recognizes that the conversion efficiency from suppressed accretion to feedback heating may be less than unity, motivating the introduction of a rescaling factor.

\section{Results}
\label{S:results}

\begin{figure*}
    \centering
    \includegraphics[width=\linewidth]{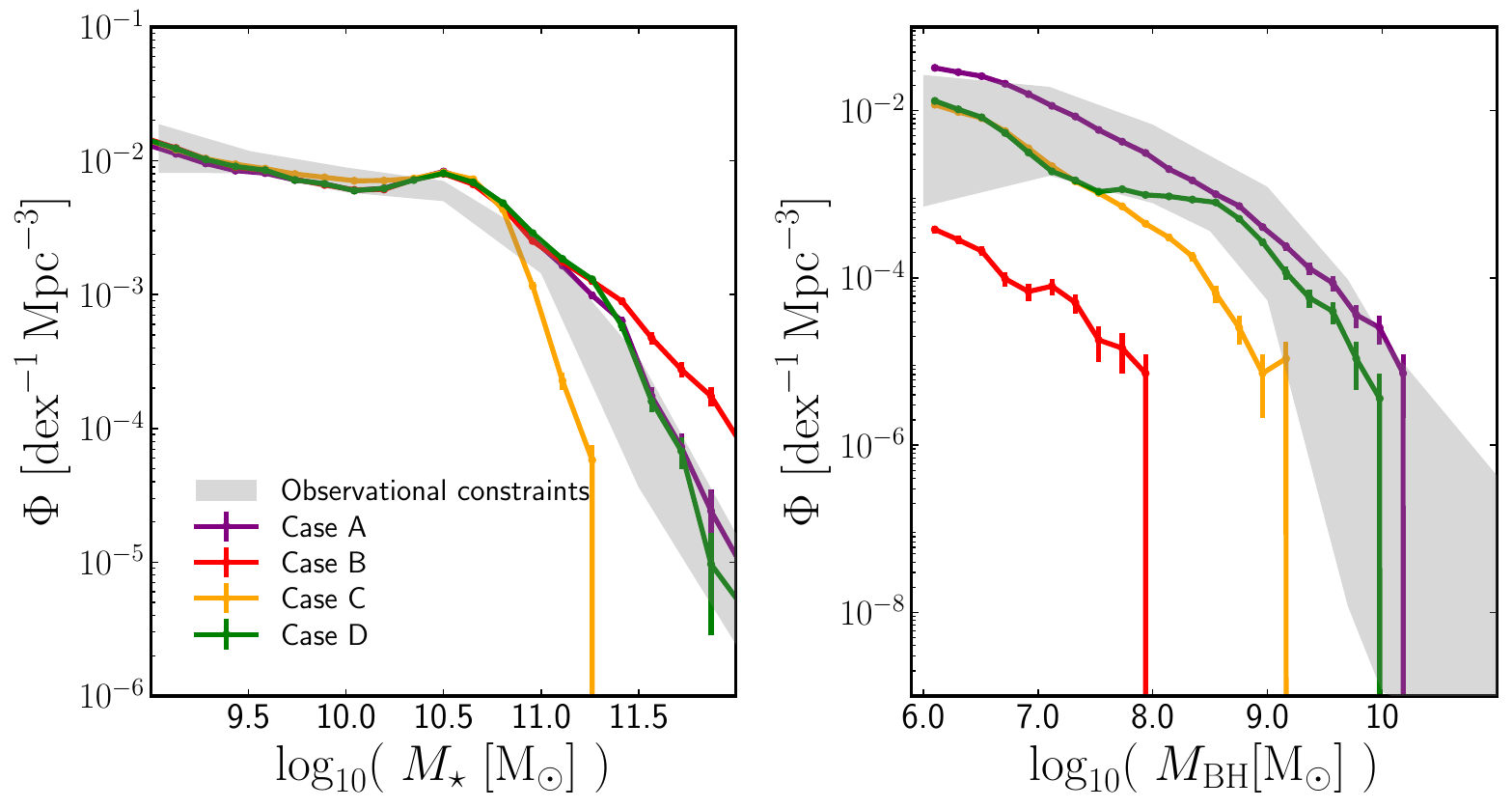}
    \caption{The stellar mass function (SMF, Left) and black hole mass function (BHMF, Right) at $z=0$ for fiducial {\sc Dark Sage} ({\bf Case A}, purple), suppressed BH accretion ({\bf Case B}, red), sub-Eddington suppressed BH accretion ({\bf Case C}, orange), and rescaled AGN suppression ({\bf Case D}, green). The grey shaded regions show the observational constrains for the SMF \citep{Baldry2008, Bernardi2013} and the BHMF \citep[e.g. see][]{PorrasValverde2025MassAssembly} in the local Universe. Among our suppression models ({\bf Cases B}, {\bf C}, {\bf D}), only the rescaling suppression model ({\bf Case D}) matches both the SMF and BHMF simultaneously.}
    \label{fig:SMFBHMF_z0}
\end{figure*}

Figure \ref{fig:SMFBHMF_z0} shows the effect of including BH growth suppression following the results of \citet{Cho+2023,Cho+2024}. The observational constraints shown in the left panel are from \cite{Baldry2008} and \cite{Bernardi2013}. The constraints in the right panel are based on inferred BHMFs using the continuity equation \citep{MerloniHeinz2008}, semiempirical methods with the {\sc TRINITY} model \citep{ZhangHaowen2023}, stellar mass-bulge mass relation \citep{Ramakrishnan2023} and \citet{Hernandez-Yevenes+2024}, stellar mass constraints \citep{Bernardi2013, Pesce2021, Liepold2024, Burke2025}, the velocity dispersion function \citep{Natarajan2009, Bernardi2010, Sato-Polito+2023}. For more detailed comparisons and further discussion, see \citet{PorrasValverde2025MassAssembly}. 

The fiducial {\sc Dark Sage} ({\bf Case A}) is calibrated to match the SMF from observations \citep{Bernardi2013}, and this in turn predicts a BHMF that also falls within the observational constraints. In contrast, we do not tune our suppression models ({\bf Cases B}, {\bf C} and {\bf D}) to match observations. The success of these models should be judged by their ability to reproduce agreement with the observed SMF and BHMF while incorporating the fundamental physics of magnetized accretion flows.

When all BH accretion channels are suppressed by the factor $\kappa_{\rm cho}$ ({\bf Case B}), the massive end of the SMF is no longer consistent with observations; however, re-tuning the free parameters $\epsilon_{\rm R}$ or $\epsilon_{\rm Q}$ can bring it into agreement. As shown in Equations~\ref{eq:radio_mode_mheat} and \ref{eq:quasarmodefeedback}, these parameters only modify the amount of gas heated by AGN feedback, which is directly related to the hot and cold gas reservoirs of galaxies. \textit{Radio-mode} feedback depends on $\dot{M}_{\rm BH}$, which is reduced by 2-3 orders of magnitude given the suppression. Adjusting the free parameters effectively ``rescales" the strength of the feedback, confining the impact of suppression only to Event Horizon scales. In contrast, the BHMF is so strongly suppressed that no tuning of free parameters can reproduce the observations. 

Suppressing BH accretion only in the sub-Eddington regime ({\bf Case C}) strongly suppresses the growth of massive galaxies. When black holes transition from sub-Eddington to $f_{\rm Edd}>3\times 10^{-3}$ accretion, $\kappa_{\rm hot}$ and $\kappa_{\rm cold}$ are set to 1. Therefore, the AGN heating rate increases dramatically. Since $\kappa_{\rm hot}$ and $\kappa_{\rm cold}$ are fixed to unity to avoid compounding suppression effects, this transition results in several orders of magnitude more efficient heating, which in turn reduces the availability of cold gas in the ISM. 

In {\sc Dark Sage}, most of the BH growth at $z>3$ comes from $f_{\rm Edd}>3\times 10^{-3}$ \citep{PorrasValverde2025MassAssembly}, so the lack of AGN feedback in this regime severely alters the evolution of the SMF. It is well understood that AGN feedback plays a significant role in producing the observed local SMF at the massive end, where massive black holes typically accrete 
at sub-Eddington rates\citep{2006MNRAS.370..645B, Croton2006, Somerville+2008, 2012MNRAS.422.2816B, Somerville2015, Croton2016, 2024ApJ...976..148P}. Therefore, for {\bf Case D}, $\epsilon_{\rm R}$ and $\epsilon_{\rm Q}$ are rescaled whenever $f_{\rm Edd}>3\times 10^{-3}$ to ensure match with the SMF. This adjustment therefore restores the simultaneous match to observations of both the local SMF and BHMF. {\bf Case D} is thus able to match the observations in the right panel in \autoref{fig:SMFBHMF_z0} almost as well as our fiducial observationally best matched {\bf Case A}.

As expected, none of the BH suppression models strongly impact the SMF at low masses, below $M_{\rm \star} < 10^{10.5}\, \mathrm{M}_{\odot}$ (left panel in \autoref{fig:SMFBHMF_z0}). At these lower stellar masses, it is supernova feedback that dominates and tunes the SMF rather than AGN feedback.

\begin{figure*}
    \centering
    \includegraphics[width=\linewidth]{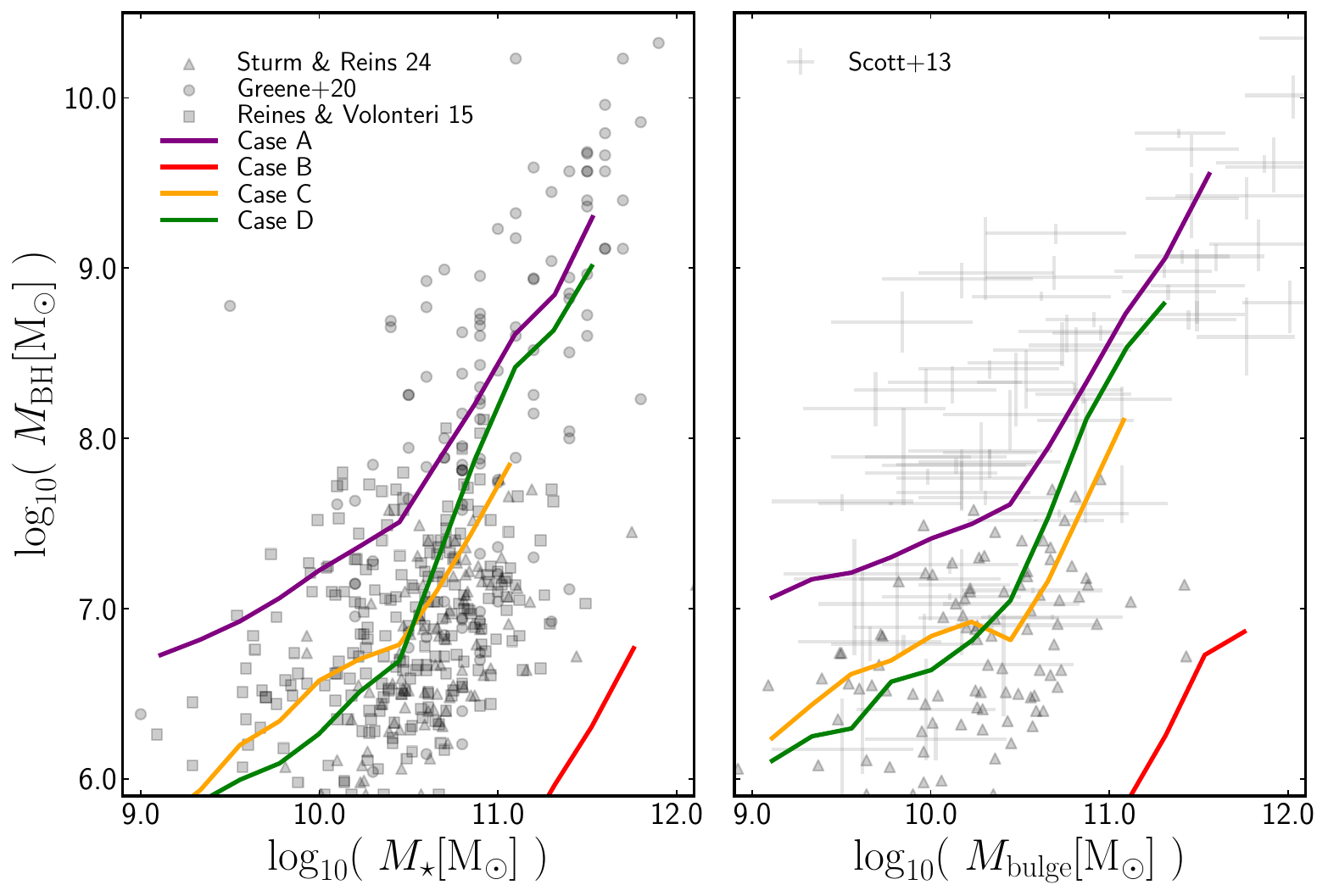}
    \caption{The $M_{\rm BH}-M_{\star}$ (left) and $M_{\rm BH}-M_{\rm bulge}$ (right) relation at z=0 for all models. We include observations from \citet{Scott2013THEGALAXIES}, \citet{Reines&Volonteri2015}, \citet{Greene2020}, and \citet{SturmReines2024}. When BH growth is entirely suppressed ({\bf Case B}), the resulting mean values are systematically reduced, yielding predictions that are completely inconsistent with observations. The other models agree qualitatively with the data.}
    \label{fig:MbhMbulge_z0}
\end{figure*}

Figure~\ref{fig:MbhMbulge_z0} shows that {\bf Cases C} and {\bf D} remain consistent with the observed $M_{\rm BH}-M_{\star}$ and $M_{\rm BH}-M_{\rm bulge}$ relations throughout the mass range, while {\bf Case A} (the fiducial model) overpredicts BH masses at the low-mass end by almost an order of magnitude. Not surprisingly, {\bf Case B}, in which all accretion modes are drastically suppressed, results in the production of significantly undermassive BHs and fails to reproduce the observed relation. As shown in Figure~\ref{fig:SMFBHMF_z0}, {\bf Case B} provides a reasonable match to the observed SMF but severely underestimates the BHMF, with a BH deficit below $M_{\rm BH} < 10^{8}\, M_{\odot}$ that shifts the normalization of the scaling relation downward. Although {\bf Case C} reproduces the BH scaling relations well, it fails to produce the most massive galaxies in the SMF. Of the scenarios studied here, only {\bf Case D} simultaneously matches the observed SMF, BHMF, and BH scaling relations, achieving agreement with observations comparable to or slightly better than the fiducial {\bf Case A}.

\begin{figure*}
    \centering
    \includegraphics[width=\linewidth]{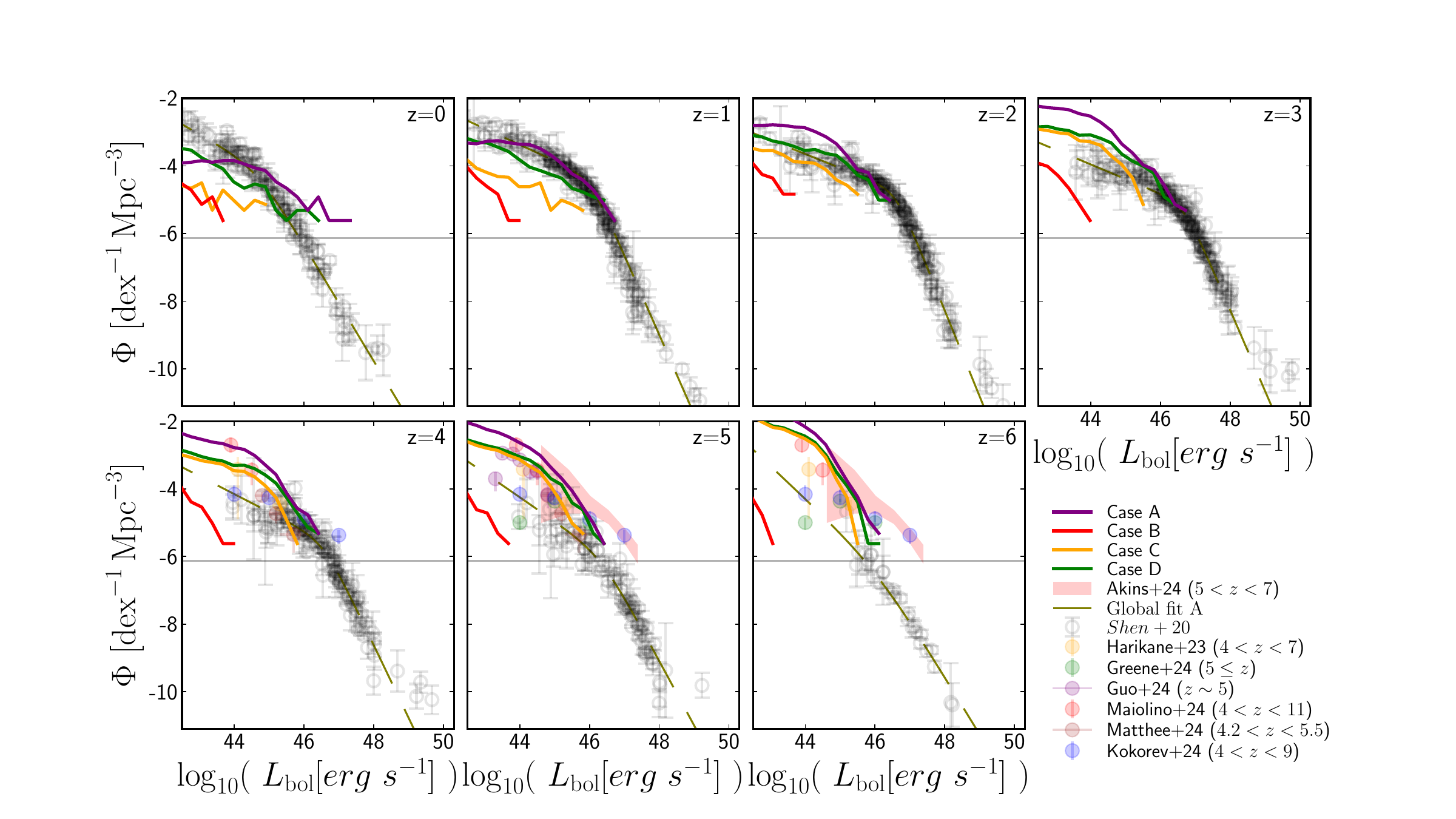}
    \caption{The bolometric AGN luminosity function across redshifts $z=0-6$ for the four models under consideration: {\bf Case A} (fiducial Dark Sage), {\bf Case B}, {\bf Case C}, {\bf Case D}. The comparison includes observational data from \citet{Shen2020AGNbol} (shown in grey), which span multiple wavebands including rest-frame infrared, B-band, ultraviolet, soft and hard X-ray, along with comprehensive bolometric, dust, and extinction corrections. \citet{Shen2020AGNbol} gives the Global fit A line, here represented by the dashed olive line. At $z=5-6$, the shaded red region indicates the inferred bolometric luminosity under the assumption that the "little red dots" identified by JWST are AGNs \citep{2024arXiv240610341A, Pacucci-Narayan2024LRD}. Additional JWST observations are included from \citet{2023ApJ...959...39H}, \citet{2024ApJ...964...39G}, \citet{2024arXiv240919205G}, \citet{2024A&A...691A.145M}, \citet{2024ApJ...963..129M}, and \citet{2024ApJ...968...38K}, each plotted at their corresponding redshift intervals. Most models show better agreement with JWST results at high redshift ($z>3$), transitioning to closer alignment with \citet{Shen2020AGNbol} at lower redshifts. Among the three tested models, {\bf Case D} gives the best match to data, performing almost as well as {\bf Case A}, and {\bf Case B} performs worst.}
    \label{fig:AGNbol_z0_z6}
\end{figure*}

Figure \ref{fig:AGNbol_z0_z6} shows the AGN bolometric luminosity function for all models compared with observations. In $z=6$, there is a close agreement between the fiducial {\sc Dark Sage} model and all the BH suppression models implemented here with observations bar {\bf Case B}. In fact, {\bf Case B} does not produce a substantial population of BHs with bolometric luminosities that are well within our target range. All other BH suppression models begin to diverge from the fiducial case at lower redshifts. Below $z<6$, the suppressed models show increasingly significant deviations from the fiducial {\sc Dark Sage} model, systematically resulting in lower AGN number densities. This trend arises because the fraction of sources accreting at sub-Eddington rates increases at lower redshifts, therefore, enhancing the effectiveness of BH suppression mechanisms. At all redshifts and a fixed bolometric luminosity, {\bf Case C} consistently produces slightly lower BH number densities compared to {\bf Case D}. This discrepancy in number densities becomes more pronounced at lower redshifts. Given the hierarchical nature of all structure assembly in the Universe, the most massive BHs are predominantly found at lower redshifts. Since AGN feedback in our models depends on BH mass, this feedback process regulates the availability of cold gas for subsequent BH growth. As demonstrated in Figures \ref{fig:SMFBHMF_z0} and \ref{fig:MbhMbulge_z0}, {\bf Case C} systematically under-produces massive BHs due to stronger AGN feedback in massive galaxies. Consequently, at $z<6$, this model begins to show significant deviations in number densities, as BH growth becomes increasingly suppressed following the transition from the high redshifts when Eddington ratios are higher $f_{\rm Edd}>3\times 10^{-3}$ phase.

\begin{figure*}
    \centering
    \includegraphics[width=\linewidth]{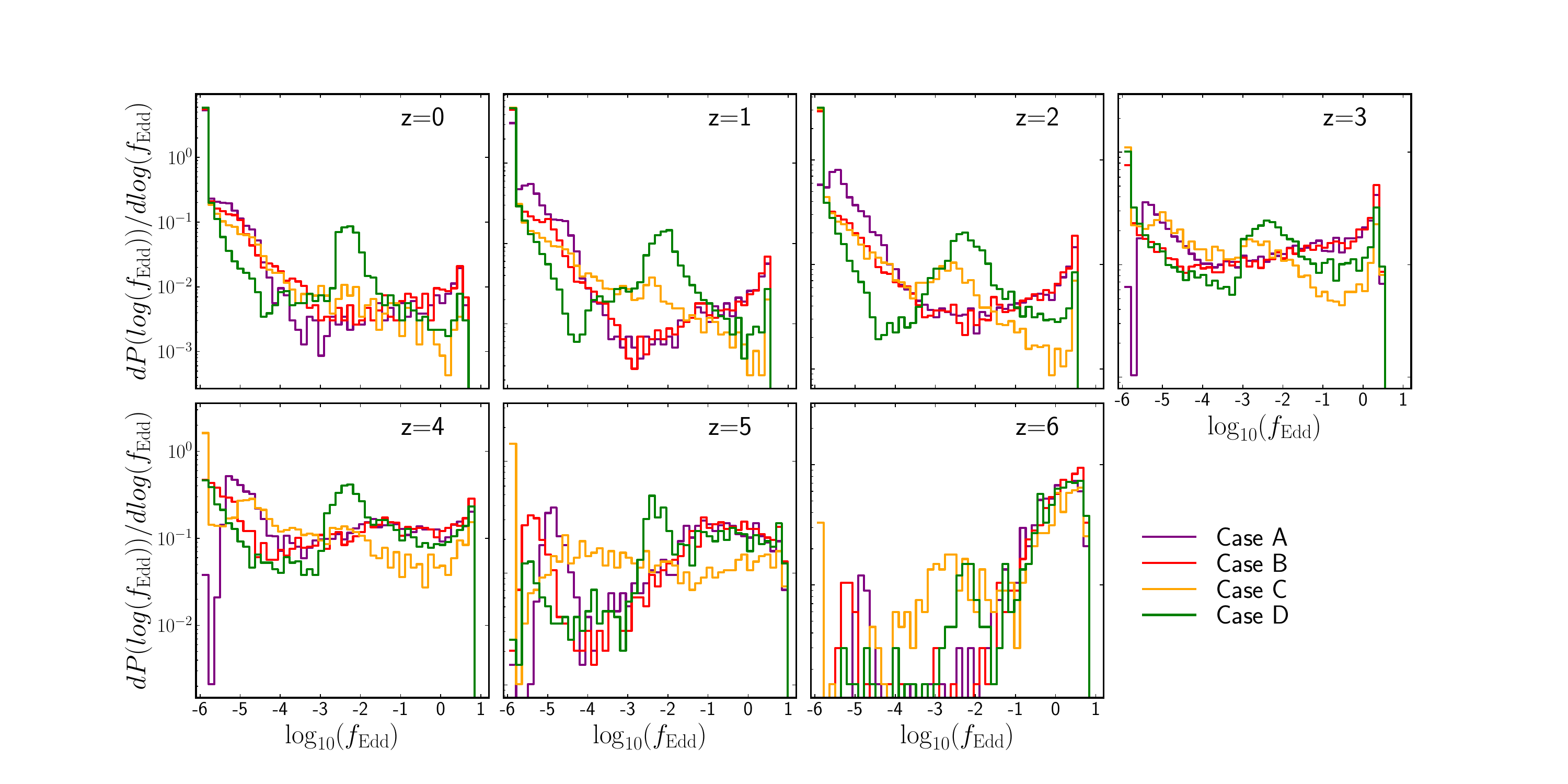}
    \caption{Histogram of the Eddington ratio distributions across redshifts $z=0-6$ for the four models under consideration: {\bf Case A} (fiducial Dark Sage), {\bf Case B}, {\bf Case C}, {\bf Case D}. All models converge at $z=6$ but show significant differences at lower redshifts.}
    \label{fig:Eddratiohist_z0_z6}
\end{figure*}

At $z>3$, the {\sc Dark Sage} models successfully reproduce the high number densities of luminous BHs observed by JWST shown in colored circles. As illustrated in Figure \ref{fig:Eddratiohist_z0_z6}, most of these sources in {\sc Dark Sage} are accreting in the super-Eddington regime. Unlike hot gas accretion (which is capped at $\dot{M}_{\rm Edd}$), cold gas accretion proceeds unrestricted at rates set by gas inflows from mergers and disk instabilities, naturally allowing $f_{\rm Edd} > 1$ to exist without additional physics. Below $z<4$, we observe a transitional phase where the fiducial {\sc Dark Sage} model and all black hole feedback suppressed models begin to overlap with the observations from \citet{Shen2020AGNbol}. Within this redshift range, super-Eddington accretion is no longer the dominant growth mode. Between $z=0-1$, {\bf Case A} and {\bf Case D} are in best agreement with \citet{Shen2020AGNbol}. This convergence primarily reflects the evolution of the underlying black hole mass function. As the BHMF builds up over cosmic time, more black holes reach masses where their bolometric luminosities fall within the observed range. 

Figure \ref{fig:Eddratiohist_z0_z6} reveals that all {\sc Dark Sage} BHs have significant super-Eddington accretion at $z=6$ despite the inclusion of our BH suppression models. Although {\bf Case B} suppresses all BH growth, the reduced AGN heating rate, as a result of the reduced accretion rate (approximately an order of magnitude lower than the fiducial case) preserves more cold gas, which, in turn, increases the available gas reservoir available for BH accretion. Interestingly, at these epochs, our BH suppression models prove ineffective against large volumes of cold gas channeled to the central BH. This effect is particularly pronounced at $z=6$ when the cold gas merger-triggered accretion channel dominates BH growth through minor galaxy mergers \citep{PorrasValverde2025MassAssembly}\footnote{\citet{PorrasValverde2025MassAssembly} shows that low-mass BHs living in low-mass halos undergo significant accretion at $f_{\rm Edd}>3\times 10^{-3}$}. 

Although these super-Eddington episodes are highly efficient at driving BHs to masses in excess of $10^6 \Msun$, their growth eventually stalls. Host galaxies subsequently accumulate stellar mass at rates exceeding BH growth, resulting in reducing the normalization of the $M_{\rm BH}-M_{\star}$ relation (figure \ref{fig:MbhMbulge_z0}) at lower redshifts. Although {\bf Case B} successfully reproduces the $f_{\rm Edd}$ distribution of the fiducial model for $f_{\rm Edd}>3\times 10^{-3}$ across all redshifts, the asymmetry in mass growth rates between galaxies and their central BHs ultimately lead to systematic underproduction of massive BHs in this model. 

In {\bf Case C}, prescribing $\epsilon_{\rm R}=\epsilon_{\rm Q}$ = 1 for sources accreting at $f_{\rm Edd}>3\times 10^{-3}$ produces significantly stronger feedback that depletes the galactic cold gas reservoir through enhanced heating. Despite this enhanced heating, the Eddington ratio distribution remains comparable to that of sources in the other models. This occurs because {\bf Case C} underproduces massive BHs by an order of magnitude compared to {\bf Cases A} and {\bf D}. Since AGN feedback scales with BH mass, the total integrated feedback in {\bf Case C} is actually weaker than {\bf Cases A} and {\bf D}, allowing substantial cold gas to remain available for accretion. For {\bf Case D}, we observe an emerging peak at $f_{\rm Edd}\sim 10^{-2}$ that develops as a result of $\epsilon_{\rm rescale}$ within the heating prescription. The sub-Eddington regime at $z<5$, in {\bf Case D} also shows a shift toward lower Eddington ratios compared to {\bf Case A} in \autoref{fig:Eddratiohist_z0_z6}.

\begin{figure}
    \centering
    \includegraphics[width=\linewidth]{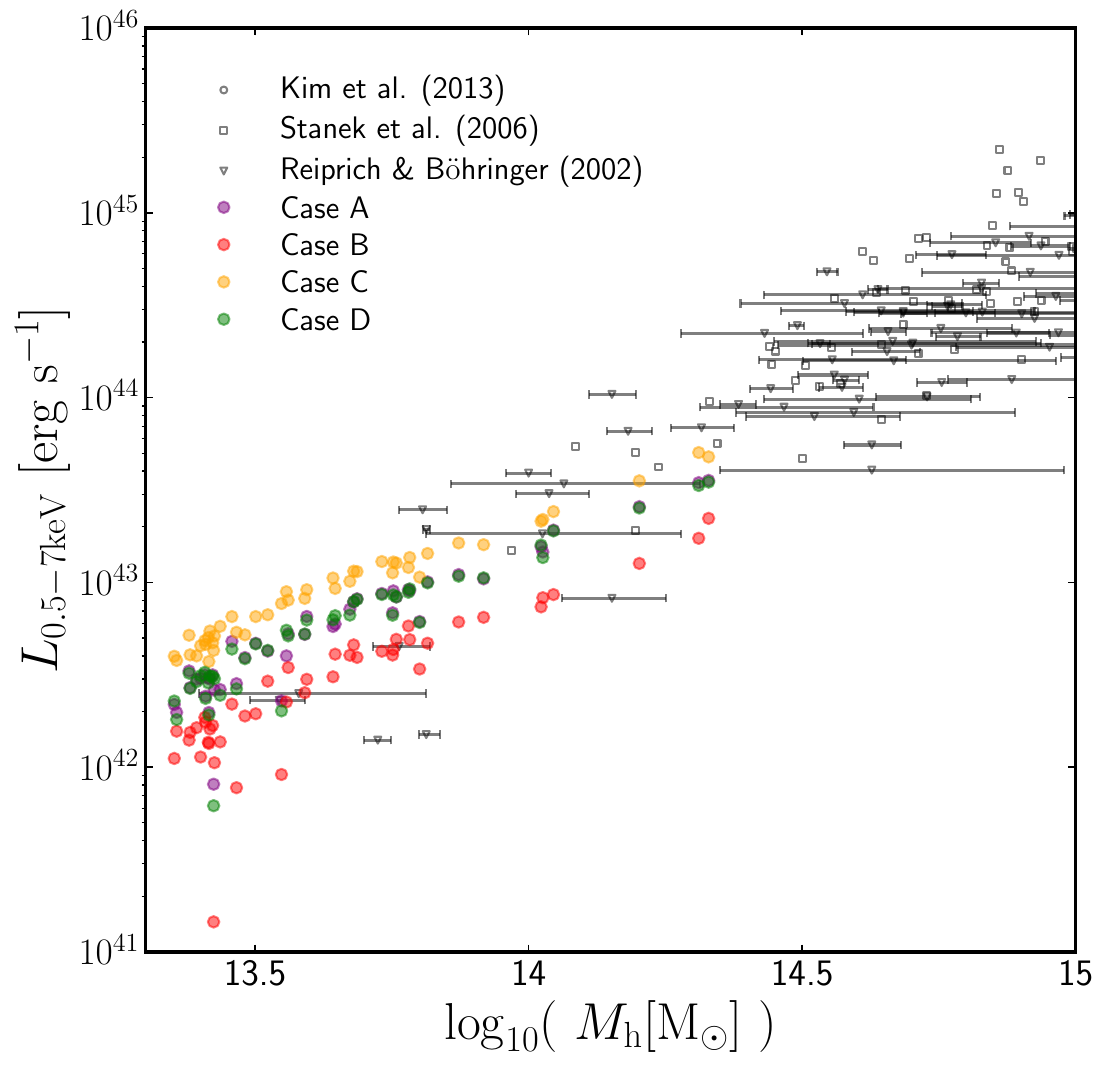}
    \caption{The X-ray luminosity in the 0.5-7 keV band as a function of halo mass for the four models under consideration: {\bf Case A} (fiducial Dark Sage), {\bf Case B}, {\bf Case C}, {\bf Case D} at $z\sim0$. Black circles, squares, and inverted triangles represent observations from \citet{Kim2013ApJ}, \citet{Stanek2006}, and \citet{ReiprichBohringer2002}, respectively. Our models fall within the observational scatter and are consistent with the observed trend. Our models do not extend to halos above $10^{14.5}\,M_\odot$ due to the finite volume of the {\sc TNG100-1} simulation box.}
    \label{fig:xraylum_z0}
\end{figure}

Figure \ref{fig:xraylum_z0} shows all halos with virial temperatures greater than 0.5 keV. We computed $L_{\rm 0.5-7\,keV}$ using the same method as in \cite{Ressler+2018}, in which the cooling curve is calculated  for the photospheric solar abundances \citep{Lodders+2003}, using the spectral analysis code SPEX \citep{Kaastra+1996} in the same way as in \cite{Schure+2009} and scaled according to the fractions of hydrogen, helium and metal mass with the assumption that the metallicity is 0.3 solar. We assumed a beta-profile gas distribution ($\beta=0.5$, and the gas halo concentration $R_{\rm vir}/r_{rm gas}=10$) normalized to the total hot gas mass at the virial temperature. 
Our function $L_{\rm 0.5-7\,keV}$ depends on the mass of hot gas, $T_{\rm vir}$, and $R_{\rm vir}$ of each halo. 

All models fall within the observational scatter and demonstrate good agreement with the data. However, the trends reveal the distinct impact of our feedback prescriptions on the hot gas reservoir. {\bf Case C} has the highest X-ray luminosities at fixed halo mass because having $\epsilon_{\rm R}=\epsilon_{\rm Q}=1$ for $f_{\rm Edd}>3\times10^{-3}$ converts all accreted mass into feedback energy, maximizing the heated gas mass in the hot reservoir. {\bf Case B} shows the  lowest X-ray luminosities because suppressing all accretion modes severely limits BH growth, resulting in both undermassive BHs and reduced heating. {\bf Case D} lies intermediate between {\bf Cases B} and {\bf C}. The rescaled coupling efficiency $\epsilon_{\rm rescale}$ reduces the total energy deposited into the hot gas reservoir, producing moderate X-ray luminosities. As discussed in Section \ref{S:methods}, while \textit{radio-mode} feedback heats the hot gas reservoir, \textit{quasar-mode} feedback can unbind and expel hot gas beyond the virial radius. However, massive halos with deeper gravitational potentials effectively retain their hot gas even with varying feedback efficiencies, such that the X-ray-emitting gas reservoir remains largely unchanged. This explains why our models reproduce the observed X-ray luminosities despite producing vastly different SMFs, particularly at the massive end. 

Figure \ref{fig:xraylum_z0} shows no halos below $M_{\rm h} < 10^{13.4}\, \mathrm{M}_{\odot}$ due to the applied cut in virial temperature of 0.5 keV. A limitation of the {\sc Dark Sage} SAM is its assumption of a single-temperature hot gas reservoir, which is then directly mapped to the virial temperature of the halo. The underlying dark matter simulation, {\sc TNG300-1}, contains halos with virial temperatures predominantly peaking at $10^{5}\, K$, which falls below the temperature range where our cooling functions are most effective. Without the 0.5 keV virial temperature threshold, {\sc Dark Sage} systematically under-predicts X-ray luminosities by several orders of magnitude. Furthermore, hot gas in low-mass halos is more susceptible to ejection by \textit{quasar-mode} feedback, further reducing the X-ray emission.

\section{Discussion}
\label{S:discussion}

The vast spatial scales between the Event Horizon and standard Bondi accretion prescriptions used in semi-analytic models and cosmological hydrodynamic simulations are challenging to resolve. The multi-zone GRMHD simulations developed by \citet{Cho+2023, Cho+2024} reveal that strong magnetic fields accumulating on horizon scales dramatically reduce accretion rates observed compared to classical Bondi theory and power mild energy transport to galactic scales, most likely due to reconnection-driven convection. We apply the BH accretion suppression factor spanning multiple BH and halo mass scales, and show how these effects propagate to influence BH and galaxy evolution.

Universal suppression of BH growth by several orders of magnitude significantly affects the massive end of local SMF ({\bf Case B}). Since BHs cannot exceed $M_{\rm BH} \sim 10^{8}\, \mathrm{M}_{\odot}$ and AGN feedback is directly coupled to $\dot{M}_{\rm BH}$, enhanced cold gas availability promotes increased star formation, resulting in enhanced galaxy growth. A more physically motivated approach, following \citet{Cho+2023, Cho+2024}, restricts the suppression to low-luminosity regimes at $f_{\rm Edd}<3\times 10^{-3}$ ({\bf Case C}), resulting in modest improvements in the local BHMF compared to the most aggressive suppression scenario. However, this approach severely curtails the growth of the galaxy at the massive end. For BHs that accumulate mass at $f_{\rm Edd}>3\times 10^{-3}$, neither accretion nor suppression of the AGN feedback operate to alter the SMF and the BHMF. Therefore, these galaxies hosting highly accreting BHs convert the same amount of accreted mass into AGN-heated gas, which is then eventually deposited into the cold gas reservoir. This enhanced heating dramatically depletes the cold gas supply, effectively quenching star formation in massive galaxies. Naturally, rescaling AGN feedback suppression, especially for $f_{\rm Edd}>3\times 10^{-3}$ sources, results in a simultaneous match to local SMF and BHMF observations ({\bf Case D}). 

Our accretion flow model in {\bf Case C} and {\bf Case D} has an abrupt transition between magnetized hot accretion and a thin disk at $f_{\rm Edd}=3\times 10^{-3}$.  In reality, this transition is likely to be more gradual, and the value at which this transition occurs is uncertain to an order of magnitude \citep[e.g.,][]{Yuan&Narayan2014,Cho&Narayan2022}.  Observationally, a transition at an Eddington ratio of $10^{-2}$ is supported both by SED changes in the AGN population \citep{Trump+2011} and changing-look AGN \citep{Ricci&Trakhtenbrot2023,Jana+2025}. Our value of $f_{\rm Edd}=3\times 10^{-3}$ is selected to be consistent with {\sc Serotina}, a semi-analytic model that tracks the evolution of BH spin self-consistently, where a larger threshold at $f_{\rm Edd}=3\times 10^{-2}$ was found to lead to potentially excessive spin-down from jets compared to X-ray reflection spectroscopy observations of AGN \citep{Ricarte+2025}.  

Lowering the $f_{\rm Edd}$ threshold, at which $\kappa_{\rm cho}$ is applied, by an order of magnitude effectively produces a closer match to the fiducial {\sc Dark Sage} BH properties analyzed in this paper. This effect is stronger at low redshift ($z<4$) due to the increased number of BHs accreting at sub-Eddington rates. Conversely, increasing the threshold by one or two orders of magnitude further suppresses BH growth, causing the model to more closely resemble {\bf Case B}. At $z=6$, we find little variation when changing the threshold, given the high fraction of BHs accreting at $f_{\rm Edd}>3\times 10^{-3}$.

Bridging Event Horizon physics and galactic-scale processes directly influence the BH-galaxy coevolution. Using the standard Bondi framework adopted by state-of-the-art cosmological hydrodynamic simulations {\sc ASTRID} and {\sc IllustrisTNG}, \citet{Dattathri2025} highlights that while {\sc ASTRID} maintains a relatively constant BH accretion rate-to-star formation rate (BHAR/SFR) ratio across both redshift and stellar mass, reflecting its non-evolving $M_{\rm BH}-M_{\star}$ relation, the {\sc IllustrisTNG} presents increasing BHAR/SFR ratios with both stellar mass and decreasing redshift, consistent with the upward evolution of its $M_{\rm BH}-M_{\star}$ relation over cosmic time. In comparison, {\sc Dark Sage} fiducial run on Millennium merger trees \citep{Springel2005Nat}, a model comparable to {\bf Case A}, shows significantly elevated BHAR/SFR ratios at $z>2$ that remain relatively constant during this epoch and exceed those found in both {\sc ASTRID} and {\sc IllustrisTNG} at similar redshifts. In {\sc Dark Sage}, high BHAR/SFR ratios arise from the dominance of merger-triggered BH accretion, which allows BHs to grow faster than their host galaxies during this epoch \citep{PorrasValverde2025MassAssembly}. By $z<2$, AGN feedback balances star formation with accretion rates \citep{Dattathri2025}. 

\citet[submitted]{Su+2025d} implemented the Bondi suppression framework from \citet{Cho+2023, Cho+2024} in cosmological zoom-in simulations. Tracking the mass assembly of SMBHs and their host galaxies in $\sim10^{14}\,M_\odot$ halos, they find that BH seeds with $M_{\rm BH} \lesssim 10^7\,M_\odot$ cannot grow through hot gas accretion alone regardless of feedback strength, with median accretion rates reaching only $10^{-2}$--$10^{-3}$ of the Bondi rate. For more massive BH seeds ($M_{\rm BH} \gtrsim 3\times10^7\,M_\odot$), a tension emerges between growth and regulation: low feedback efficiency ($\eta \lesssim 0.02$, corresponding to low spin $a_* \sim 0$) allows BH growth but fails to quench star formation, while high efficiency ($\eta \geq 0.3$, high spin $a_* \geq 0.9$) effectively quenches star formation but suppresses BH accretion. This work highlights that rapid cold gas accretion channels are essential for BH growth below $\sim10^7\,M_\odot$, while earlier and more massive seeding combined with mergers can potentially reconcile the local $M_{\rm BH}$--$M_{\star}$ relation independent of hot gas accretion.

The question of whether star formation and BH growth were all more efficient at higher redshifts, coupled with strong or weak feedback, emerges as central to reconciling JWST observations with theoretical predictions \citep{Koudmani2019, Sharma2020, Su+2023, Boylan-Kolchin2023Nature, Endsley2023, Dekel2023, Li2024, Yung2024, Scholtz2025}. Our case models suggest that the answer is nuanced: while super-Eddington accretion provides rapid BH growth at high redshift, the concurrent need to regulate star formation creates tension between achieving massive BH populations and maintaining a realistic galaxy mass function. Our bolometric AGN luminosity function shows a transition, where below $z<4$, some models converge with \citet{Shen2020AGNbol} observations. This transition corresponds to the decline of BHs accreting at $f_{\rm Edd}>3\times 10^{-3}$ and the increasing importance of feedback-regulated growth, suggesting that evolutionary efficiency indeed varies significantly across cosmic time.

Our results underscore the extreme sensitivity of galaxy mass assembly to AGN feedback. Although we can recover the locally observed SMF through the modification of $\epsilon$ (see equations \ref{eq:radio_mode_mheat} and \ref{eq:quasarmodefeedback}) when applying our BH suppression factor, the corresponding BHMFs become indifferent to the modifications. This finding reveals the careful balance required between accretion and feedback processes, where modifications to Event Horizon scale physics have cascading effects on galactic-scale observables. The transition between sub-Eddington and $f_{\rm Edd}>3\times 10^{-3}$ proves particularly critical, with heating rates increasing by several orders of magnitude when $\kappa_{\rm cho}$ transitions from suppressed to unity values. This dramatic feedback enhancement severely limits cold gas availability, showing how implementation details in feedback prescriptions can fundamentally alter evolutionary pathways. 

Our parameterized approach through $\epsilon_{\rm R}$ and $\epsilon_{\rm Q}$ implicitly assumes constant coupling and feedback efficiencies, yet the underlying physics demands that the feedback strength vary with both the BH spin evolution and the accretion state. BH spin fundamentally determines jet efficiency through the Blandford-Znajek mechanism \citep{Blandford1977, Tchekhovskoy2011}, with rapidly spinning BHs extracting rotational energy to power jets at efficiencies potentially exceeding 100\% \citep{Tchekhovskoy2011, Narayan2022}. 

\section{Conclusions}
\label{S:conclusions}

In this paper, we implement a BH accretion suppression factor derived from GRMHD simulations into the semi-analytic model {\sc Dark Sage}. In summary, our findings

\begin{itemize}
    \item Suppressing all BH accretion ({\bf Case B}) increases the massive end of the SMF by indirectly reducing the total feedback energy injected. In contrast, suppressing only BHs accreting at $f_{\rm Edd}<3\times 10^{-3}$ and rescaling ({\bf Case D}) the coupling efficiency (the efficiency associated with AGN feedback) provides a simultaneous match to both the local SMF and BHMF.
    \item When comparing the AGN bolometric luminosities in Figure \ref{fig:AGNbol_z0_z6}, all suppression models except {\bf Case B} agree well with the fiducial model at $z=6$, but diverge toward lower redshifts as the fraction of sub-Eddington accretors increases, leading to systematically lower BH number densities. This effect is least noticeable in {\bf Case D}.
    \item At $z>3$, {\sc Dark Sage} models successfully reproduce the high number densities of luminous BHs observed by JWST, with most sources accreting super-Eddington. Below $z<4$, all models converge with \citet{Shen2020AGNbol} as accretion above $f_{\rm Edd}>3\times 10^{-3}$ decreases.
\end{itemize}

Our implementation of $\kappa_{\rm cho}$ reveals that magnetic fields cannot universally suppress accretion rates across all environments and epochs, as evidenced by the super-Eddington growth required to match JWST observations at $z>6$. \citet{Cho+2023, Cho+2024} demonstrate that magnetic suppression is most effective in hot, low-density flows characteristic of \textit{radio-mode} accretion. Physically, BH accretion appears to transition from the magnetically-arrested disk (MAD) to the standard and naturally rotating (SANE) regime. Some GRMHD simulations suggest that this transition is neither sharp nor universal \citep{Narayan2012, Ryan2017, Mizuno2022, Anantua2024}. Intermediate states arise where magnetic fields provide partial suppression of accretion \citep{Begelman2022, Raha2025arXiv}.  This poses a fundamental challenge for semi-analytic models, which may require accretion and/or feedback suppression factors that must simultaneously explain both the low accretion rates in quiescent systems and the rapid super-Eddington growth necessary to grow the massive black holes observed by JWST at $z>6$.  

\begin{acknowledgments}
AJPV gratefully acknowledges the overlapped support of a Postdoctoral fellowship from the Heising-Simons foundation and the National Science Foundation Astronomy \& Astrophysics Postdoctoral Fellowship. This material is based on work supported by the National Science Foundation under Award Number 2502826. AJPV thanks Adam R. H. Stevens for the fruitful discussions about the project. We used computational facilities from the Yale Center for Research Computing (YCRC). The literature reviews for this work were made using the NASA’s Astrophysics Data System. P.N., A.R., K.S., H.C., R.N. and B.P. acknowledge support from the Gordon and Betty Moore Foundation and the John Templeton Foundation that fund the Black Hole Initiative (BHI) at Harvard University.
\end{acknowledgments}

\bibliography{bridgingscales_bibfile}{}
\bibliographystyle{aasjournal}

\appendix

\section{Distribution of $\kappa_{\rm cho}$ values}
\label{app:kappa_cho_distribution}

\begin{figure*}
    \centering
    \includegraphics[width=\linewidth]{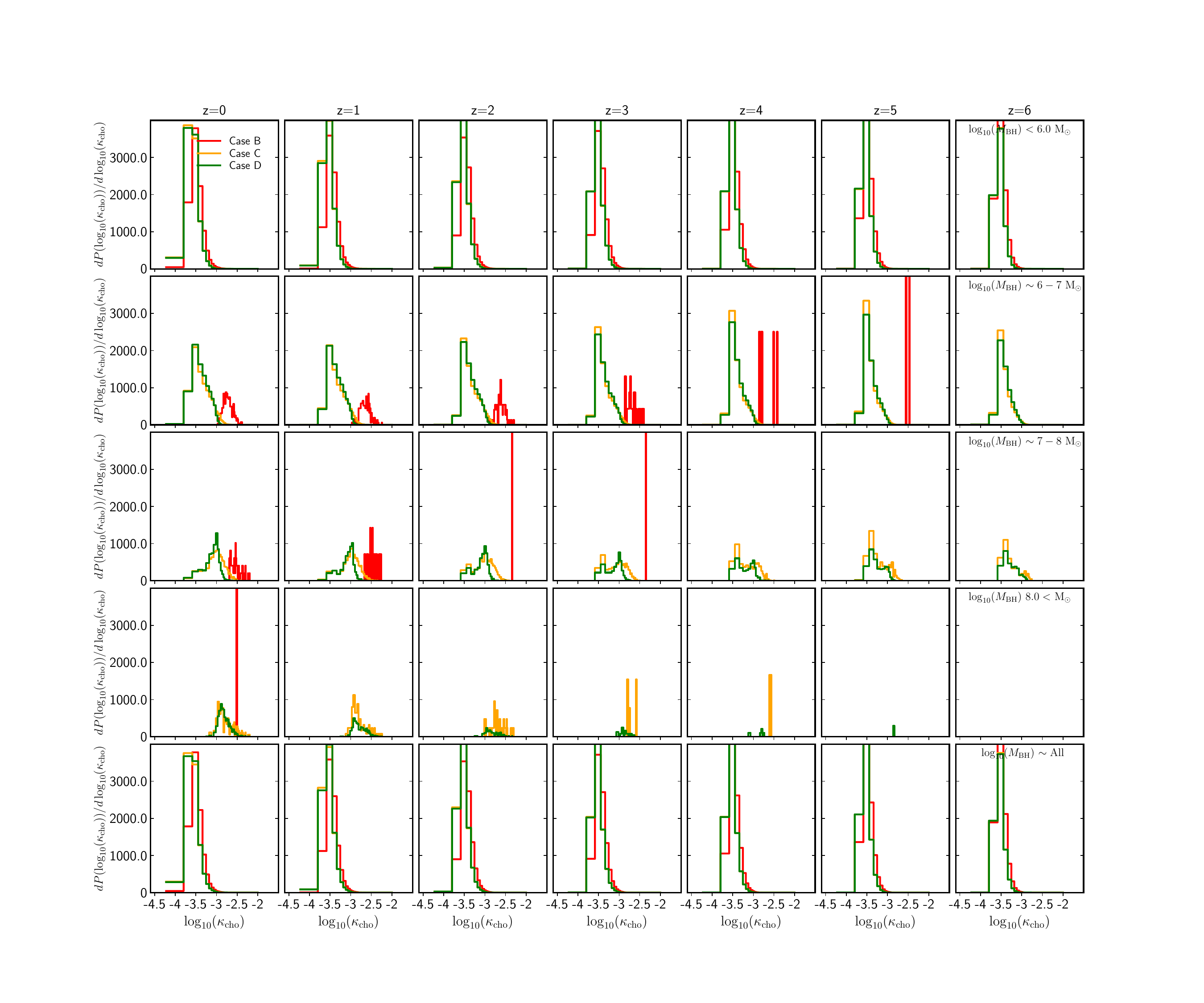}
    \caption{Distribution of $\kappa_{\rm cho}$ at $z=0$--6 for different BH mass bins. All models show a consistent peak at $\kappa_{\rm cho} \sim 10^{-3}$ across cosmic time.}
    \label{fig:kappachohist_BHbins_z0_z6}
\end{figure*}

To understand how the Bondi suppression factor $\kappa_{\rm cho}$ varies across BH mass and cosmic time, we examine its distribution for 
{\bf Cases B}, {\bf C}, and {\bf D} at $z=0$--6 in Figure \ref{fig:kappachohist_BHbins_z0_z6}. Across all models and redshifts, the distribution consistently peaks at $\kappa_{\rm cho} \sim 10^{-3.5}$, which is driven by low-mass BHs ($M_{\rm BH}<10^6\,M_\odot$). For the most massive black holes ($M_{\rm BH} > 10^7\,M_\odot$), all models 
show broader distributions extending to larger $\kappa_{\rm cho}$ values, particularly at lower redshifts. This trend reflects the increasing diversity of accretion modes as these systems transition from predominantly super-Eddington accretion at high redshift to more varied sub-Eddington regimes at late times.

Using kernel-weighted averages of density and temperature within an adaptive smoothing radius, \citet{Su+2025d} compute $\kappa_{\rm cho}$ and track its temporal evolution alongside the effective gas temperature and scaled Bondi radius ($r_{\rm Bondi}/r_g$). Both our semi-analytic model results and the \citet{Su+2025d} cosmological zoom-in simulations find that $\kappa_{\rm cho}$ consistently lies in the range $10^{-3}$--$10^{-2}$, with median values of $1$--$3\times10^{-3}$. \citet{Su+2025d} find that more massive black holes produce broader distributions of $\kappa$ due to stronger feedback episodes temporarily heating the surrounding gas, while simultaneously encountering denser gas in deeper potential wells, resulting in a lower median $\kappa_{\rm cho}$.

The broader high-$\kappa_{\rm cho}$ tail in our models at low redshift reflects the increasing fraction of BHs accreting across a diversity of growth channels, whereas \citet{Su+2025d}'s broader distributions at high $\kappa_{\rm cho}$ reflect local temperature variations around individual BHs and episodic feedback heating rather than accretion mode diversity. Note, that \citet{Su+2025d} only models M87*-type BHs while here we apply $\kappa_{\rm cho}$ across a wider range of BH masses.

\end{CJK}
\end{document}